\documentstyle[aps,preprint,epsf]{revtex}

\author{ 
Rava da Silveira$^{\ast}$, Sahraoui Chaieb$^{\dagger}$ and L. Mahadevan$^{\dagger}$\\ 
{\footnotesize
{\it $^{\ast}$Department of Physics, $^{\dagger}$Department of Mechanical Engineering}}\\
{\footnotesize
{\it Massachusetts Institute of Technology, 77 Mass Ave., Cambridge, MA 02139, U.S.A.}}}

\title{Rippling Instability of a Collapsing Bubble
}

\begin{document}

\maketitle


\begin{abstract}
When a bubble of air rises to the top of a highly viscous liquid, it forms a
dome-shaped protuberance on the free surface. Unlike a soap bubble, it
bursts so slowly as to collapse under its own weight simultaneously, and
folds into a striking wavy structure. This rippling effect occurs in fact
for both elastic and viscous sheets, and a theory for its onset is
formulated. The growth of the corrugation is governed by the competition
between gravitational and bending (shearing) forces and is exhibited for a
range of densities, stiffnesses (viscosities), and sizes -- a result which
arises less from dynamics than from geometry, suggesting a wide validity. A
quantitative expression for the number of ripples is presented, together
with experimental results which are in agreement with the theoretical
predictions.
\end{abstract}

\newpage\ 

Every day, nature surprises us with structures and patterns of such beauty
as to fill the scientist with wonder and the artist with envy. 
In the present paper, we address an instability which turns a hemispherical,
smooth, liquid bubble into a striking wrinkled structure, first observed by
Debr\'egeas, de Gennes, and Brochard-Wyart \cite{debregeas1}. In their
experiment, 0.1 to 10 cm$^3$ of air injected into a highly viscous liquid ($%
\eta \sim 10^3$ Pa$\cdot $s) rises to the free surface, imprisoned in a
hemispherical bubble of thickness $t\sim 1-10~\mu $m. If the latter is
punctured at its apex by a needle, surface tension drives the rapid
expansion of a circular opening. The retraction velocity soon (after about $%
10-30$ ms) saturates to a constant, owing to the high viscous resistance. In
the meantime, the air flow through the hole equilibrates the pressure
difference, allowing the bubble to collapse under its own weight. As it
deflates, an {\it instability} appears: the fluid sheet folds into a wavy
structure, with radial ripples that break the original axisymmetry. In the
absence of a detailed theory, Ref.~\cite{debregeas1} proposes a scaling
estimate for the number of ripples $n^{*}\sim \left( \mu gR^3/K\right)
^{1/2} $, where $\mu $ is the mass of the film per unit area, $g$ the
gravitational acceleration, $R$ the radius of the hole, and $K$ an effective
bending rigidity of the sheet which was assumed to be elastic during the
early stages of the rippling.

The rippling results from the competition between compression, bending, and
gravity. Each fluid element tends to fall under its own weight, but
experiences a viscous resistance from its neighborhood. If the bubble were
to collapse in a uniform, symmetric way, it would occupy a progressively
reduced area, leading to an in-plane compression which would require forces
that far exceed the scale set by gravity. Instead, the film deforms in a
nearly inextensional fashion by undergoing pure bending. Equivalently, for a
given (gravitational) force, the relative time scale associated with
stretching is much larger than that for bending, and the surface therefore
corrugates over short times, before eventually relaxing into a uniform,
thicker membrane.

This instability is reminiscent of buckling phenomena \cite{brush},
originally studied in the context of elastic rods but occurring also in the
creeping flows of viscous liquid filaments (a striking everyday example
being the coiling of a stream of honey when it reaches a toast \cite
{taylor,mahadevan}). For an elastic rod, buckling occurs at the longest
possible wavelength in order to minimize the bending energy. In the bubble
problem, however, gravity plays a distinctive role in determining the
configuration. For a given amplitude, bending still favors large scale
deformations, while gravitational energy is minimized for an almost flat
sheet with as many tiny ripples as possible; the optimal wavelength results
from a compromise between the two. Such an argument, however, does not fully
characterize the effect. Unlike the above examples, here the system under
consideration is a {\it curved two-dimensional} sheet, and the
associated geometry constrains the rippling both qualitatively and
quantitatively, as we shall see.

The instability occurs both an elastic (solid) film and a viscous (liquid)
one. The elastic case corresponds to a convex curved shell with an apical hole of
radius $R$ that is allowed to collapse under its own weight. In the viscous case,
an additional complication arises because the radius of the hole changes
during the rippling. Following a short initial transient, the hole grows
steadily at a rate $v\sim \sigma /\eta $ resulting from the balance of
surface tension ($\sigma \approx 20~$mN/m) and viscous stress. It thus takes
a time $\tau \sim \eta t/\sigma $ for the opening radius to increase by $t$.
During this time, the liquid acquires a velocity $V\sim g\tau $ due to
gravity, larger than $v$ by a factor $V/v\sim 10^7$. Even if the liquid is
viscoelastic, so that the retraction velocity is enhanced by a factor $R/t$ (%
$\sim 10-10^4$) \cite{debregeas2,debregeas1}, the hole radius remains
essentially constant while the instability occurs ({\it cf}. Fig.~1). We may
therefore treat the hole radius $R$ as a given parameter in the theory \cite
{surface-tension}.

Although the bubble has the geometry of a sphere before collapsing, it is
quite flattened by the time the ripples appear (Fig.~1(B)). For simplicity,
we consider the unperturbed configuration to be a shallow cone, of slope $%
\alpha \ll 1$, described by its height above the surface, 
\begin{equation}
\label{cone}h=\alpha \left( r _0-r \right) , 
\end{equation}
where $r $ is the cylindrical radial coordinate and $r _0$ the radius of the
base. Any deformation of $h$ introduced by the rippling may be written,
without loss of generality, as 
\begin{equation}
\label{perturbation}h+\delta h=\alpha \left( r _0-r \right) -\delta \alpha
\left( r \right) +\sum_{n\geq 1}\left\{ \delta \beta _n^{\left( 1\right)
}\left( r \right) \cos \left( n\theta \right) +\delta \beta _n^{\left(
2\right) }\left( r \right) \sin \left( n\theta \right) \right\} , 
\end{equation}
where $\theta $ is the azimuthal angle. The perturbation $\delta \alpha $
represents a uniform ($n$ independent) flattening accompanying the growth of
ripples of amplitude $\delta \beta _n^{\left( i\right) }$, and a crucial
step consists in understanding their form and interdependence. In the case
of a thin elastic (viscous) sheet, the two primary modes of deformation are
in-plane stretching (shearing) and out-of-plane bending. A generic
deformation of an elastic cone (made of a material with Young modulus $Y$),
of amplitude $\zeta $ on a scale $\ell $, requires stretching forces (per
unit surface) of order $Yt\zeta^2 /\ell ^2$, but significantly smaller
stretching forces (per unit surface) of order $Yt^3\zeta^2 /\ell ^4$ \cite
{elasticity}, so that for a given external force, here gravity,
inextensional deformations are greatly preferred \cite{curved}. In the case
of a highly viscous sheet, forces arise from velocity gradients, thus
introducing a dynamical element in the problem. However, their dependence on 
$t$ and $\ell $ (essentially due to the variation of the strain across the
film) is similar, so that inextensional deformations are again largely
favored if $t\ll l\approx r _0/n^{*}$. (This condition is satisfied if the
selected number of ripples $n^{*}$ is small compared to $10^3$, which is the
case as we shall see below.) Equivalently, for a given loading, the time
scale corresponding to bending is smaller than that for stretching by a
factor $\left( t/l\right) ^2$ \cite{buckmaster,howell}. Thus, at the onset
of the instability, perturbations of the cone must preserve its metric. This
requirement translates into the constraints\cite{exact} 
\begin{equation}
\label{inextensibility-1}\delta \beta _n^{\left( i\right) }\left( r \right)
=\delta \beta _n^{\left( i\right) }\times r +\delta \beta _n^{\left(
i\right) \prime }, 
\end{equation}
(where $\delta \beta _n^{\left( i\right) }$ and $\delta \beta _n^{\left(
i\right) \prime }$ are constants) and 
\begin{equation}
\label{inextensibility-2}4\alpha \delta \alpha \left( r \right) =\sum_{n\geq
1,i}\left\{ \left( n^2-1\right) \delta \beta _n^{\left( i\right) 2}\left( r
-r _0\right) +n^2\delta \beta _n^{\left( i\right) \prime 2}\left( \frac 1r
-\frac 1{r _0}\right) \right\} . 
\end{equation}

In the following, we elucidate the elastic (solid) case before extending our
treatment to the viscous (liquid) case. The energy functional of a perturbed
elastic cone reads

\begin{eqnarray}
E\left[ h+\delta h\right]  & = & \int_{
\text{cone}}d\left( \text{surface}\right) \times \left( \text{gravitational
potential energy }+\text{ bending potential energy}\right) \nonumber \\  & = & 
\int_R^{r
_0}r dr \int_0^{2\pi }d\theta \sqrt{1+\left[ {\bf \nabla }\left(
h+\delta h\right) \right] ^2}\left\{ \mu g\left( h+\delta h\right) 
+\frac{K}2\left(
\nabla ^2 \delta h\right) ^2\right\} .
\label{energy} \end{eqnarray}
where $K=Yt^3/12\left( 1-\nu ^2\right) $ is the
rigidity and $\nu $ the Poisson ratio. Only bending elastic energy appears
in $E$, since we have confined ourselves to the class of inextensible
deformations.

If the elastic cone is attached to the plane on which it rests so that $%
\delta h\left( r =r _0\right) =0$, Eq.~(\ref{inextensibility-1}) yields $%
\delta \beta _n^{\left( i\right) \prime }=-\delta \beta _n^{\left( i\right)
}r _0$. On substituting Eqs.~(\ref{perturbation}--\ref{inextensibility-2})
into Eq.~(\ref{energy}) we then obtain, to lowest order in the perturbation,

\begin{eqnarray}
\delta E & \equiv  & E\left[ h+\delta h\right] -E\left[ h\right] \nonumber \\  
& = & \frac \pi 2K\cdot f \left( \frac{r _0}R\right)\cdot 
\sum_{n\geq 1,i}\left( \delta \beta^{\left(i\right)}
 _n\right) ^2\left\{g\left( \frac{r _0}R,
\gamma R^3 \right) +2\left[\varphi\left( \frac{r _0}R
\right) -\gamma R^3\psi
\left( \frac{r _0}R\right) \right] n^2+n^4\right\} \nonumber \\  & \equiv  & 
\sum_{n\geq 1}\delta E_n.
\label{variation} \end{eqnarray}
$\gamma ^{-1/3}=\left( \alpha K/\mu g\right) ^{1/3}$ is an intrinsic length
scale arising from the competition between gravity and bending elasticity 
\cite{functions}.

Each mode contributes an amount $\delta E_n$ to the change in energy, and
rippling occurs if $\delta E_n<0$ for some integer. In general, $\delta
E_n<0 $ for a range of different $n$'s; the most negative variation
corresponds to the maximally growing perturbation, and thus sets the
wavelength of the instability. The formulation also yields a ``threshold
condition'' $\varphi \left( r _0/R\right) <\gamma R^3\psi \left( r
_0/R\right) $ for the occurrence of rippling. This condition involves the
three independent quantities $\gamma $, $r _0$, and $R$, and may be
translated into three corresponding statements. ({\it i}) Rippling is
suppressed if $\gamma <\gamma _{\text{c}}\left( r _0,R\right) =R^{-3}\varphi
/\psi $, {\it i.e.} if the cone is too light or too rigid. ({\it ii})
Similarly, no rippling occurs if the hole, or equivalently the cone, is too
small, $r _0<r _{0 \text{c}}\left( \gamma ,R\right) $. Azimuthal continuity
requires the wavelength of the deformation to be at most of order $r _0$,
resulting in a forbidding bending cost if $r _0$ becomes small compared to
the intrinsic (energetically determined) scale $\gamma ^{-1/3}$. ({\it iii})
Finally, the threshold depends, quite unexpectedly, on the {\it ratio} $r
_0/R$. The dependence of the symmetric ($n=0$) mode on the radial coordinate 
$r $ is different from that of the rippling ($n\neq 0$) modes, so that the
high elastic cost cannot be justified by gravitational gain anymore if the
hole is reduced beyond a critical size. Minimizing $\delta E$ in Eq.~(\ref
{variation}) yields the selected number of ripples as 
\begin{equation}
\label{scaling}n^{*}=\text{Int}\sqrt{\frac{\mu gR^3}K\cdot \frac 1\alpha
\psi \left( \frac{r _0}R\right) -\varphi \left( \frac{r _0}R\right) }, 
\end{equation}
where Int$\left\{ x\right\} $ is the integer closest to $x$. This relation
improves on the estimate of Ref.~\cite{debregeas1}, where the authors
consider the short time elastic behavior, and establishes its domain of
validity.

For an elastic (solid) sheet, the rippling phenomenon is of an essentially
static nature; upon increasing, say, the mass of the sheet, the equilibrium
configuration is shifted from symmetric to rippled. Approaching the problem from a
dynamical perspective by considering the elastic forces and torques rather
than the corresponding energies results in an evolution equation $\pi \mu
P\left( r _0,R\right) \cdot d^2\left( \delta \beta _n^{\left( i\right)
}\right) /dt^2=-\delta E/\delta \beta _n^{\left( i\right) }$ for each mode.
Here $P$ is a polynomial function independent of $n$, so that the
energetically optimal mode, with number $n^{*}$, is indeed the fastest
growing one. In the case of a viscous liquid, the effect is intrinsically
dynamical: bending occurs only on short times, while the equilibrium
configuration is ultimately reached by a slow thickening. Nevertheless, the
motion of a viscous film satisfies a formulation close to that of an elastic
sheet, as can be shown by integrating the Stokes equation through the
thickness \cite{howell,silveira}. Indeed, it is easy to see that bending
results from a torque $\frac{\eta t^3}{4\left( 1-\nu ^2\right) }\times d($%
curvature$)/dt$ analogous to an elastic torque $K\times ($curvature$)$, so
that a highly viscous film may be described by an effective bending modulus $%
K_l=\eta t^3/3\tau $ ($\nu =1/2$ for an incompressible medium), where $\tau $
is a time scale associated with the falling velocity. Thus, all the
conclusions of the stability analysis for the elastic cone, and in
particular the expression for the number of ripples (Eq.~(\ref{scaling})),
may be transposed to the case of the bubble modulo a certain time scale
related to the gravity-induced velocity of the fluid. Comparing the nascent
ripples' amplitude to the film thickness yields an estimate of this time
scale as $(t/g)^{1/2}$ \cite{variable-n}.

In order to check our results against experiment, we visualized the bursting
of silicone oil bubbles. Once the bubble is punctured with a sharp needle,
its evolution is followed using a high-speed camera capable of recording up
to 1000 frames per second. The resulting video is then analyzed to determine
the radius $r _0$ of the bubble, the hole size $R$ at which the ripples are
first observed, along with the number of ripples $n^{*}$. Since the hole
expands very fast at first, $R$ is much larger than $R_{\text{c}}$ by the
time the bubble begins to collapse. To compare the experiments with the
theory, in which $R$ enters as a parameter, the latter is measured at the
onset of the instability for each given size of the bubble. The quantitative
measurements are compared to the theoretical predictions for the dependence
of $n^{*}$ on the bubble size on Fig.~2. On a more qualitative level, the
experiments show a suppression of the instability for small bubbles, in
agreement with the threshold conditions above.

We conclude with a discussion of possible refinements of the theory and
their relation to the geometric nature of the problem. A more complete
theory would incorporate a (flattened) hemisphere as the initial condition,
rather than a cone. Also, due to the progressive drainage of the liquid, the
thickness $t$ acquires a dependence on $r $ (and time). This in turn implies
non-uniform rigidity $K\left( r \right) $ and mass $\mu \left( r \right) $,
leading to functions $f$, $g$, $\varphi $, $\psi $, and $P$ of a more
complicated form. On a more fundamental level, all these aspects should be
addressed in terms of the coupled hydrodynamics of the slow viscous (liquid)
flow and the rapid air flow; the corresponding calculations will be reported
elsewhere \cite{silveira}. Yet, the strong geometrical constraints involved
in the problem are suggestive of the robustness of the results. The question
we have answered is akin to that of applying a curved surface unto a flat
one in the most economical way, a problem which has taxed cartographers for
many centuries and lies at the birth of differential geometry. It is also
somewhat of an inverse counterpart to the problem of fitting a flat sheet to
a three dimensional landscape, which has been studied in various contexts 
\cite{kantor,lobkovsky,cerda}, and is an issue that still vexes fashion
designers. The relevance of the geometrical constraints is manifest, for
example, in the strong dependence of the rippling on the size of the
opening, which is closely related to a well-known theorem due to Gauss \cite
{spivak}, Jellett \cite{jellett}, and others, according to which (loosely
put) a closed surface cannot be bent without being stretched, while an open
surface can be bent inextensionally. Similarly, we find that a smaller hole
implies a relatively stiffer bubble, and hampers the rippling. While the
detailed form of the functions $\varphi $ and $\psi $ arise from the
physical constraints and dynamics imposed by the forces and various boundary
conditions, the essence is in the geometry.

\newpage\ 

\smallskip
FIG. 1. Stroboscopic images of a collapsing liquid bubble 
of size is $r_0= 1\text{cm}$ and thickness $t\approx 100\mu \text{m}$. 
The silicone oil has viscosity $\eta=10^3\text{Pa}\cdot \text{s}$, 
surface tension $\sigma=21\text{mN/m}$, and mass density $0.98\text{g/cm$^3$}$.
(A) $30 ms$ after the film is punctured by a sharp needle,
the bubble shows a retracting hole of radius $R=1.4\text{mm}$ at its apex, 
but no ripples yet.
(B) $30 ms$ later, the bubble loses its axisymmetric shape. The radius
of the hole remains essentially constant, at $R=1.6\text{mm}$, while the
ripples grow. 
The inset displays a schematic side view of the essentially conical
deflating bubble at the onset of the instability, with the important quantities
involved in the phenomenon. The extreme shallowness allows for a 
perturbative treatment in the slope $\alpha$ of the cone.



\bigskip
FIG. 2. Plot of the number of ripples $n^*$ as a function of the bubble radius
$r_0$, comparing the experimental measures (points) with the theoretical
predictions (solid lines). These data were gathered using silicone oil of
viscosity $\eta=600\text{Pa}\cdot\text{s}$ and bubbles of thickness
$t\approx30\mu\text{m}$.
The errors in the measurement of $r_0$ arise
from meniscus effects which are more important in smaller bubbles. The
bursting time elapsed up to rippling is measured to be of order of 1 to
5 times $\left(t/g\right)^{1/2}$, consistent with our proposed mechanism
for the formation of the corrugation.
For each experimental realization, the ratio 
$r_0 / R$ was measured at the onset of the instability, and the
corresponding dependence of $R$ on $r_0$ was used to get a
theoretical curve $n^*=n^*\left( r_0 \right)$. 
The thinner line displays the prediction for an elastic sheet attached to 
the plane on which it rests. The liquid, however, can in no way be clamped,
and one has to relax the boundary conditions at the base. This leads to
a vanishing of the unprimed modes ({\it{cf}}. Eq.~(\ref{inextensibility-1})),
which are unfavorable in terms of both graviational and viscous forces;
the fastest growing primed modes lead to the behavior represented by the 
thicker line.
The latter is plotted here for a slope $\alpha \approx
3^{\circ}(\approx 0.05 \text{rad})$ of the cone, consistent with our 
perturbative treatment, and in agreement with direct observation.
The dashed line represents the best fit of the scaling form
$n^* \sim \left( \mu gR^3/K\right) ^{1/2}$
\cite{debregeas1}, where $R$ is chosen as the relevant length scale. If
$R$ is replaced by $r_0$,
the above expression for $n^*$ may be closely fitted (up to an overall
multiplicative factor) to our predicted curve, showing that the size of
the bubble is the dominant length scale within the present experimental
range and conditions. This is consistent with the relaxed boundary 
conditions, which allow the ripples to be significant close to the outer edge
of the bubble ({\it{cf}}. also Fig.~1.(B)). In this way,
the ripples trade a bulk gain in gravitational and bending stresses
against a cost in stretching in a thin rim close to the outer edge. The 
increased thickness of the liquid film close to the base further emphasizes
this effect, as it reduces the difference in magnitude between a typical
stretching and a typical bending stress.
The inset shows a top view
of the fully developed ripples, from which $n^*$ is measured.



\begin{thebibliography}{99}
\bibitem{surface-tension}  While surface tension drives the expansion of the
hole, rippling is dominated by gravity and viscosity, and we neglect any
surface tension-related considerations in the remainder.
\bibitem{curved}  The fact that the unperturbed surface is curved plays an
essential role here. For a plane, bending energy dominates in the limit a
small deformation, with respect to stretching. (For more details, see Ref.~%
\cite{elasticity}.)
\bibitem{exact}  Exact inextensibility requires that each point on the
surface be displaced in the three dimensions in which the cone is imbedded:
vertical, radial, and azimuthal. Of course, the three displacements are
coupled, and, in the limit $\alpha \ll 1$, the radial and angular components
are negligibly smaller (by a factor $\alpha $) than the vertical
deformation. Also, we note that the form of these ``inextensibility
conditions'' depends on the unperturbed configuration, which, in particular,
is curved.
\bibitem{functions}  Here 
$$
\varphi \left( x\right) =\frac{2\left( x-1-\ln x\right) }{x^2-4x+3+2\ln x}%
\geq 0\text{,\quad }\psi \left( x\right) =\frac{\left( x-1\right) ^3}{%
3\left( x^2-4x+3+2\ln x\right) }\geq 0\text{\quad (for }x\in \left[ 1,\infty
\right] \text{),} 
$$
and $f\left( x\right) $, $g\left( x,y\right) $ are similar rational
functions of $x$, $\ln x$, and $y$.
\bibitem{variable-n}  As the bubble continues to fall, this time scale may
vary and so may the number of ripples. However, this last possibility is
unlikely owing to the large forces required to introduce new ripples or
remove existing ones.
\end{thebibliography}
\end{document}